\RequirePackage[2020-02-02]{latexrelease}
\documentclass[aip,pop,graphicx,reprint,superscriptaddress]{revtex4-2}

\usepackage{amsmath}
\usepackage{amsfonts}
\usepackage{graphicx}
\usepackage{amssymb}
\usepackage{bm}
\usepackage{braket}
\usepackage{tikz}

 \newcommand{\YZ}[1]{\textcolor{black}{#1}}

\begin{document}

\title{Nonlinear magnetohydrodynamic modeling of ideal ballooning modes in high-$\beta$ Wendelstein 7-X plasmas}

\author{Yao Zhou}
\affiliation{School of Physics and Astronomy, Institute of Natural Sciences, and MOE-LSC, Shanghai Jiao Tong University, Shanghai 200240, China}
\email[]{yao.zhou@sjtu.edu.cn}
\author{K.~Aleynikova}
\affiliation{Max-Planck-Institut f\"ur Plasmaphysik, 17491 Greifswald, Germany}
\author{Chang Liu}
\affiliation{State Key Laboratory of Nuclear Physics and Technology, School of Physics, Peking University, Beijing 100871, China}
\author{N.~M.~Ferraro}
\affiliation{Princeton Plasma Physics Laboratory, Princeton, New Jersey 08543, USA}

\date{\today}

\begin{abstract}
We present nonlinear magnetohydrodynamic (MHD) simulations of high-$\beta$ Wendelstein 7-X plasmas using the stellarator extension of the M3D-$C^1$ code, building on the recent work that shows benign saturation of ideal ballooning modes above the designed $\beta$ limit in the standard configuration [Y. Zhou et al, Phys. Rev. Lett. 133, 135102 (2024)]. First, we examine the results' sensitivity to the parallel thermal conductivity. It is found that while an increased parallel conductivity reduces the linear growth rate, the saturated pressure profile is barely affected. Second, we consider the dependence on the profile shape. It is shown that an equilibrium with a peaked pressure profile and lower $\beta$ is subject to more significant change than a broad profile with higher $\beta$ and a larger growth rate, suggesting that benign saturation, or nonlinear stability, is not guaranteed and not dictated by linear growth. 
Third, we study the influence of the magnetic configuration, with the equilibrium rotational transform varied by adjusting the planar coil current. With similar growth rates, similar magnitudes of profile change are found regardless of the presence of a low-order resonance, which implies that the saturation mechanism is not specific to a resonant or non-resonant mode. These results indicate that MHD stability should still be treated seriously in stellarator operation and design, for which nonlinear modeling using tools like M3D-$C^1$ can play an instrumental role.

\end{abstract}

\maketitle

\section{Introduction}
The stellarator is one of the most promising concepts for future fusion reactors, featuring unique advantages in steady-state operation by avoiding challenges associated with the maintenance and stability of the plasma current \cite{Boozer2005}. The experimental findings of the advanced Wendelstein 7-X (W7-X) stellarator have successfully demonstrated that optimized magnetic configurations can significantly improve plasma confinement, achieving record-breaking stellarator fusion-triple-product values with modest heating power \cite{Beidler2021}. Subsequently, W7-X has been equipped with water-cooled divertors and enhanced fueling and heating capabilities that enabled further progress in extended high-performance operation \cite{Grulke2025,Langenberg2024}. Ultimately, the continuous, reactor-relevant operation of W7-X will have profound implications on the viability of the stellarator concept, and its success critically depends on maintaining magnetohydrodynamic (MHD) stability at high $\beta$ (which denotes the volume-averaged ratio of plasma to magnetic pressure throughout this paper).  

The designed $\beta$-limit of W7-X is about $5\%$ based on linear MHD stability analysis \cite{Beidler1990, Grieger1992}, but stellarator plasmas have shown nonlinear stability when driven beyond linear stability thresholds in experiments \cite{Weller2006}. Recently, we reported the first nonlinear MHD simulations of pressure-driven instabilities in high-$\beta$ W7-X plasmas \cite{Zhou2024}, which were enabled by the stellarator extension of the M3D-$C^1$ code \cite{Zhou2021}.
In the standard configuration, the simulation results agree with the design analysis that ideal ballooning modes become linearly unstable when $\beta$ exceeds 5\%, but predict nonlinear saturation at benign levels with mild confinement degradation, which is much less severe than the original expectation of losing the plasma column. This suggests that W7-X may be able to operate at or slightly above the designed $\beta$-limit while avoiding major MHD events, and enhances the appeal of the stellarator approach to steady-state fusion reactors.
However, the conclusion needs to be further examined since the results can be sensitive to simulation parameters, profile shapes, magnetic configurations, etc. 

In this paper, we build on the results in Ref.~\onlinecite{Zhou2024} and present more comprehensive MHD simulation results of ideal ballooning modes in high-$\beta$ W7-X plasmas.
First, we examine the results' sensitivity to the parallel thermal conductivity, and find that while an increased parallel conductivity reduces the linear growth rate, the saturated change of the pressure profile is barely affected. Second, we consider the results' dependence on the profile shape. We show that an equilibrium with a peaked pressure profile and lower $\beta$ is subject to more significant degradation than a broad profile with higher $\beta$ and a larger growth rate, suggesting that benign saturation, or nonlinear stability, is not guaranteed and not dictated by linear growth. 
Third, we study the impact of the magnetic configuration, varying the equilibrium rotational transform by adjusting the planar coil current. We find that similar growth rates result in similar magnitudes of profile change, regardless of the presence of a low-order resonance. This implies that the saturation mechanism is not specific to a resonant or non-resonant mode. These results indicate that MHD stability should still be treated seriously in stellarator operation and design, for which nonlinear modeling using tools like M3D-$C^1$ can play an instrumental role. 

This paper is organized as follows. In Section \ref{sec:settings} we introduce the M3D-$C^1$ code and the simulation settings. In Section \ref{sec:kappa} we analyze the sensitivity of linear growth and nonlinear saturation with respect to the parallel heat conductivity. In Section \ref{sec:prof} we show that a peaked pressure profile is subject to more pronounced ballooning-induced degradation than a broad profile. In Section \ref{sec:iota} we study how the saturated state depends on the equilibrium rotational transform. Summary and discussion follow in Section \ref{sec:sum}.

\section{Simulation settings}\label{sec:settings}

M3D-$C^1$ is an open-source nonlinear MHD code designed to model the macroscopic dynamics of toroidal fusion plasmas \cite{Jardin2012,Ferraro2018}. 
For temporal discretization, M3D-$C^1$ implements a split-implicit scheme that realizes stable and efficient transport-timescale simulations \cite{Jardin2012b}. 
For spatial discretization, M3D-$C^1$ employs high-order finite elements with $C^1$ continuity in three dimensions and has been successfully extended to stellarator geometry \cite{Zhou2021}. 
Recent simulations of current-drive-induced sawtooth-like crashes in W7-X validates this new capability by showing semi-quantitative agreements with experimental results \cite{Zhou2023}.
While two-fluid and many other effects are available in M3D-$C^1$, their functionality has not yet been fully verified in stellarator geometry. 
So in this work, we solve the single-fluid extended MHD equations (in SI units)
\begin{align}
\partial_t \rho + \nabla\cdot(\rho\mathbf{v}) =& D\nabla^2(\rho-\rho_0),\label{continuity}\\
\rho(\partial_t \mathbf{v} + \mathbf{v}\cdot\nabla\mathbf{v}) =& \mathbf{j}\times\mathbf{B} - \nabla p - \nabla\cdot\mathbf{\Pi},\label{momentum}\\
\partial_t p + \mathbf{v}\cdot\nabla p +\gamma p\nabla\cdot\mathbf{v} =& (\gamma-1)[\eta ({j}^2-{j}_0^2)& \nonumber\\
&- \nabla\cdot\mathbf{q} - \mathbf{\Pi} : \nabla\mathbf{v}],&\label{energy}\\
\partial_t \mathbf{B} =&  \nabla\times[\mathbf{v}\times\mathbf{B} - \eta (\mathbf{j}-\mathbf{j}_0)],\label{induction}
\end{align}
for the mass density $\rho$, velocity $\mathbf{v}$, pressure $p$, and magnetic field $\mathbf{B}$, with the current density $ \mathbf{j}= \mu_0^{-1}\nabla\times\mathbf{B}$ and the adiabatic index $\gamma=5/3$. 
The viscous stress tensor is $\mathbf{\Pi} = -\mu(\nabla\mathbf{v}+\nabla\mathbf{v}^{\mathrm{T}}) - 2(\mu_{\mathrm{c}}-\mu)(\nabla\cdot\mathbf{v})\mathbf{I}$ and the heat flux 
$\mathbf{q} = -\kappa_\perp\nabla(T-T_0) - \kappa_\parallel\mathbf{b}\mathbf{b}\cdot\nabla T$
with $\mathbf{b} =\mathbf{B}/B$ and the temperature $T=Mp/\rho$, where $M$ is the ion mass.  

The transport coefficients used include uniform mass diffusivity $D=2.18~\mathrm{m^2/s}$, isotropic and compressible viscosities $\mu=\mu_\mathrm{c}=3.65\times10^{-7}~\mathrm{kg/(m\cdot s)}$. The ratio between the parallel and perpendicular thermal conductivities, $\kappa_{\parallel}/\kappa_{\perp}$, is varied with $\kappa_{\perp}$ set to $\kappa_{0}=2.18\times10^{20}~\mathrm{(m\cdot s)^{-1}}$ unless otherwise noted (c.f. Section \ref{sec:kappa}). 
The temperature-dependent resistivity $\eta$ is enhanced by a factor of 100 from the Spitzer resistivity and its typical value is about $10^{-6}~\mathrm{\Omega\cdot m}$ (core) to $10^{-5}~\mathrm{\Omega\cdot m}$ (edge). 
The equilibrium fields $\rho_0$, $T_0$, and $\mathbf{j}_0$ subtracted in the dissipative terms act as effective sources to sustain the equilibrium in the absence of instabilities.
The equilibrium density and temperature profiles are $\rho_0\propto p_0^{1/\gamma}$ and $T_0\propto p_0^{1-1/\gamma}$, unless otherwise noted (see Section \ref{sec:prof}).
We use 3807 reduced quintic elements in the $(R,Z)$ plane and 160 Hermite cubic elements in the toroidal direction (for a full torus).
We use a time step size of 0.573 $\mu$s to simulate a hydrogen plasma with a core number density of $1.5\times10^{20}~\mathrm{m}^{-3}$, such that the core temperature is about 5 keV (assuming equipartition). These settings are mostly the same as those in Ref.~\onlinecite{Zhou2024} except for the varying $\kappa_{\parallel}/\kappa_{\perp}$. 

We initialize fixed-boundary stellarator simulations in M3D-$C^1$ using the outputs of the widely used 3D equilibrium code VMEC \cite{Hirshman1983}, including the geometry of the flux surfaces as well as the magnetic field $\mathbf{B}_0$ and the pressure $p_0$.
The former is given in terms of a coordinate mapping, $R(s,\theta,\varphi)$ and $Z(s,\theta,\varphi)$ with $s$ being the normalized toroidal flux and $\theta$ the poloidal angle in VMEC, which is utilized to set up the non-axisymmetric computational domain {enclosed by the last closed flux surface}. (This mapping does not evolve in M3D-$C^1$ so that $s$ and $\theta$ are fixed regardless of the actual dynamics of the flux surfaces.)
The latter then provide the initial conditions and subtracted equilibrium fields in the simulations. 
{The boundary conditions are ideal on the magnetic field, no-slip on the velocity, and fixed on the density and pressure.}

\section{Sensitivity to the parallel thermal conductivity}\label{sec:kappa}
In magnetically confined plasmas, heat transport is strongly anisotropic. 
The ratio between parallel and perpendicular thermal conductivities, $\kappa_{\parallel}/\kappa_{\perp}$, can be as high as $\sim 10^8$ according to experimental measurements \cite{Holzl2009}, posing significant challenges for accurate numerical modeling. 
Explicit codes often adopt substantially reduced $\kappa_\parallel$ to avoid the formidably small time steps required by numerical stability. 
For implicit codes, resolving realistically strong anisotropy also requires carefully designed numerical methods \cite{Jardin2012b}. 
The high-order finite element method adopted by M3D-$C^1$ has been demonstrated to be effective in this aspect \cite{Jardin2004}, and this favorable feature is inherited by the stellarator extension by utilizing the geometry of the flux surfaces to set up the mapping from the logical to the physical coordinates \cite{Zhou2021}.

\begin{figure}
\centering
\includegraphics[width=0.85\columnwidth]{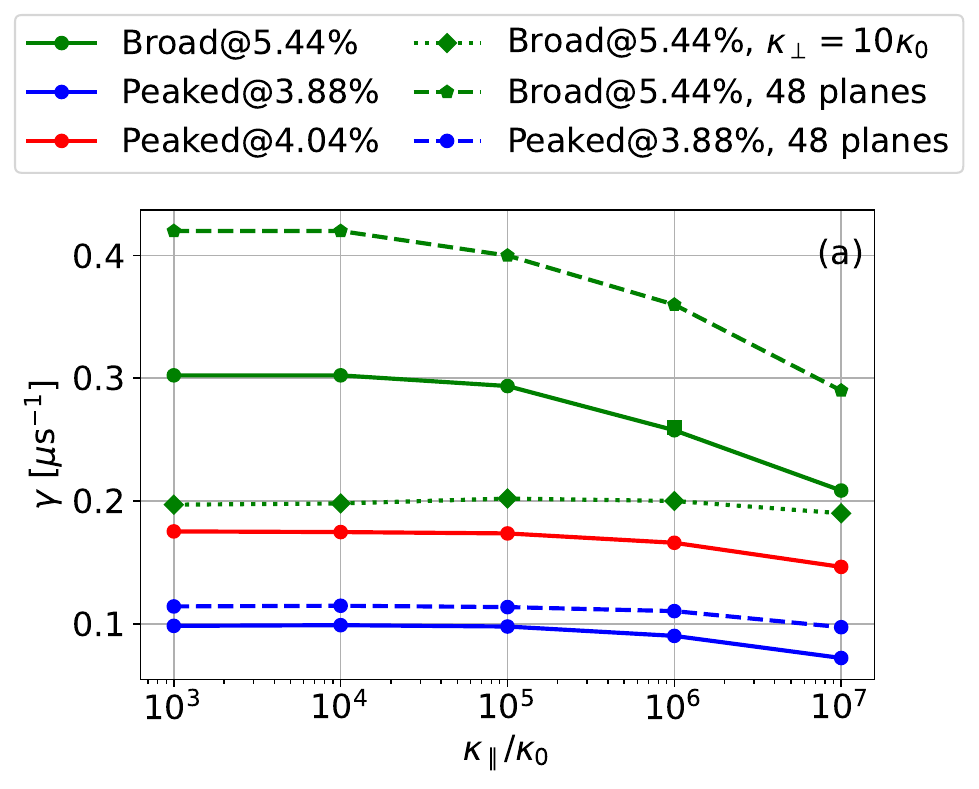}
\includegraphics[width=0.85\columnwidth]{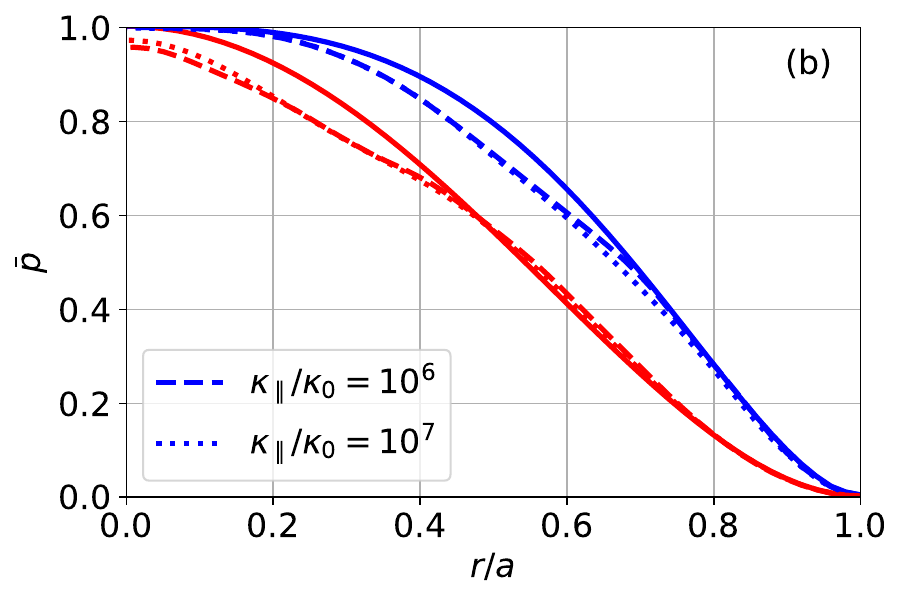}
\caption{(a) The linear growth rate $\gamma$ vs. $\kappa_{\parallel}/\kappa_{0}$ in single-field-period simulations of the standard EIM configuration using a broad pressure profile (green) and a peaked pressure profile (blue and red). The dashed curves show results obtained with higher toroidal resolution (48 planes). The dotted curve shows results obtained with $\kappa_{\perp}=10\kappa_{0}$. (b) The saturated shapes of the pressure profile $\bar{p}$ at $\theta=0$ and $\varphi=\pi$ in full-torus simulations with $\kappa_{\parallel}/\kappa_{0} = 10^6$ (dashed) and $10^7$ (dotted), compared to the respective initial shapes (solid; blue: broad profile with $\beta=5.44\%$; red: peaked profile with $\beta=3.88\%$). The normalized minor radius $r/a=\sqrt{s}$.}
\label{fig:growth}
\end{figure}

However, in practice, the value of $\kappa_{\parallel}$ is often chosen somewhat arbitrarily for convenience. 
In Ref.~\onlinecite{Zhou2024}, $\kappa_{\parallel}$ is set to $10^6\kappa_{0}$ and not varied. 
Considering that the pressure evolution can be important for ballooning modes, here we examine the sensitivity of their growth rates with respect to $\kappa_{\parallel}$ using single-field-period simulations, which model 1/5 of the torus using 32 elements in the toroidal direction. 
Here, we consider the standard `EIM' configuration of W7-X, specifically the $\beta=5.44\%$ equilibrium with a broad, parabolic-like pressure profile and zero current profile, and $B\approx 2.3~\mathrm{T}$ in the core. 
The dependence of the linear growth rate $\gamma$ versus $\kappa_{\parallel}/\kappa_{0}$ is shown by the green solid curve in Figure \ref{fig:growth}(a). 
It can be seen that when $\kappa_{\parallel}/\kappa_{0}> 10^4$, $\gamma$ decreases as $\kappa_{\parallel}$ increases, indicating that a sufficiently large, or close-to-realistic parallel heat conductivity can \YZ{significantly reduce} the linear growth of the ballooning mode in the simulations. 

To better understand this \YZ{reduction}, we increase $\kappa_{\perp}$ to $10\kappa_{0}$, and the dependence of $\gamma$ versus $\kappa_{\parallel}/\kappa_{0}$ is shown by the green dotted curve. 
The growth rates are now \YZ{much less sensitive} to the change in $\kappa_{\parallel}$, and their values are close to that with $\kappa_{\parallel}/\kappa_{0}= 10^7$ and $\kappa_{\perp}=\kappa_{0}$. This suggests that $\kappa_{\perp}$ itself can also stabilize the ballooning modes, and that the stabilizing effect of $\kappa_{\parallel}$ is partly due to numerical pollution \YZ{(extra, non-physical perpendicular transport due to the discretization errors of parallel diffusion)}. 
In addition, we increase the toroidal resolution \YZ{from 32} to 48 elements in a single period, which resolves higher-$n$ modes and mitigates numerical pollution simultaneously. 
The growth rates shown by the green dashed curve are larger and have a similar dependence on $\kappa_{\parallel}$ to those obtained with lower resolution, which indicates that $\kappa_{\parallel}$ itself also has a direct stabilizing effect on the ballooning modes that is not due to numerical pollution.

The linear growth rates' sensitivity to the numerical parameters and resolution highlights the formidable challenges in accurately resolving the nonlinear dynamics of the ballooning modes. 
Fortunately, the nonlinearly saturated states seem rather insensitive to $\kappa_{\parallel}$ and hence can serve as more meaningful observables. 
In particular, we focus on the pressure profile change induced in full-torus simulations as a key nonlinear metric of the severity of the ballooning modes. 
In Figure \ref{fig:growth}(b) the blue dashed curve shows the saturated pressure profile obtained with $\kappa_{\parallel}/\kappa_{0} = 10^6$ in Ref.~\onlinecite{Zhou2024}, which agrees well with the blue dotted curve obtained using $\kappa_{\parallel}/\kappa_{0} = 10^7$, despite the different linear growth rates. 
Therefore, the specific choice of $\kappa_{\parallel}$ used in Ref.~\onlinecite{Zhou2024} does not have a significant impact on its main conclusion, that the ballooning modes saturate at benign levels in the standard configuration. 
The results obtained using a different initial pressure profile (red curves, c.f. Section  \ref{sec:prof}) also show similar features.
In the following, we use $\kappa_{\parallel}/\kappa_{0} = 10^7$ in the simulations for better numerical stabilty.

\section{Dependence on the profile shape}\label{sec:prof}

\begin{figure}
\centering
\includegraphics[width=\columnwidth]{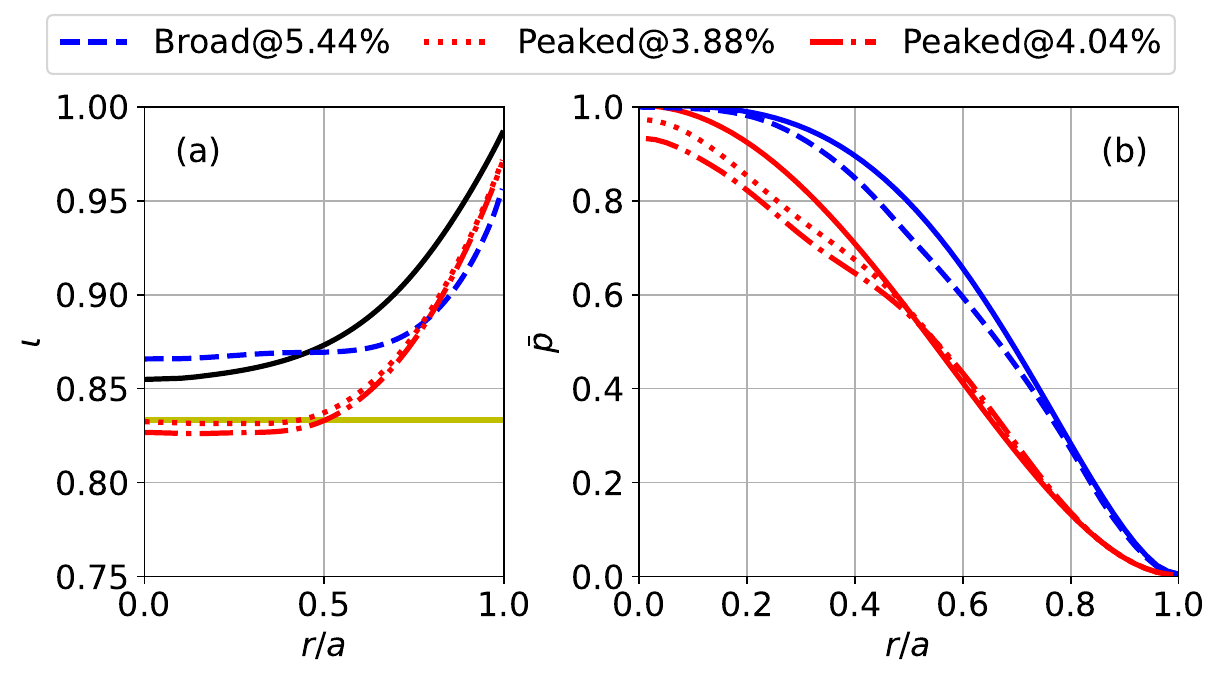}
\caption{(a) The rotational transform profiles in the simulated EIM equilibria, with the vacuum profile (black solid) shown for comparison. The yellow solid line marks the $\iota=5/6$ resonance. (b) The saturated shapes of the pressure profile in the M3D-$C^1$ simulations, with the initial shapes (solid; blue: broad profile with $\beta=5.44\%$; red: peaked profile with $\beta=3.88\%$ and $4.04\%$) shown for comparison.}
\label{fig:prof}
\end{figure}

\begin{figure*}
\centering
\includegraphics[width=\textwidth]{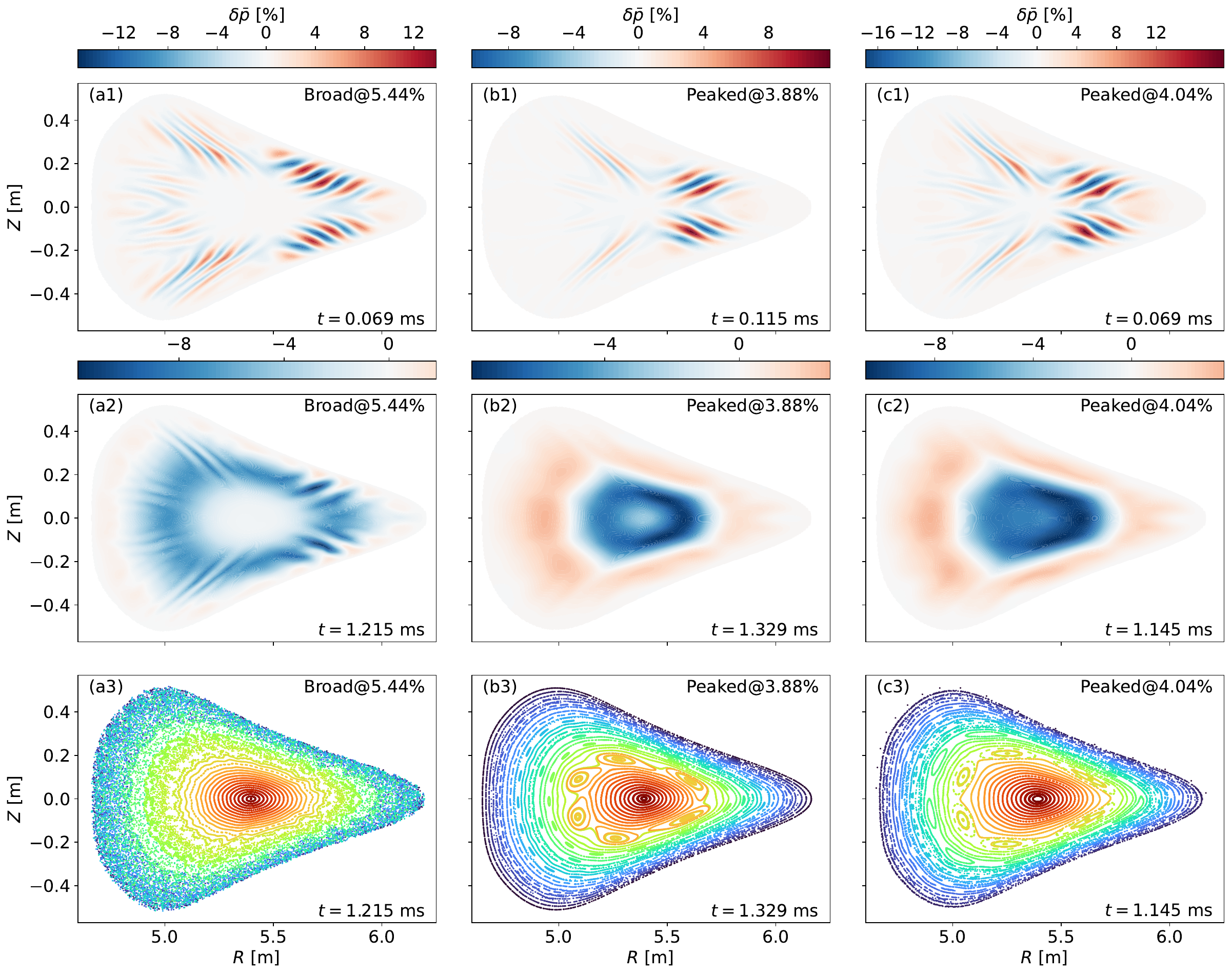}
\caption{Snapshots of the normalized pressure change in the three simulations in Figure \ref{fig:prof}: row 1 shows mode structures at the end of the linear growth phase, while row 2 shows the saturated states at the end of the simulations. Note the different scales in the color bars. Row 3 shows the Poincar\'e plots of the saturated states, where different colors label different field-lines.}
\label{fig:snap}
\end{figure*}

So far, we have focused on the broad, parabolic-like pressure profile shown by the blue solid curve in Figure \ref{fig:growth}(b), where $\bar{p}$ is the pressure normalized by its initial core value. 
This profile is favorable for achieving high volume-averaged $\beta$, and the original W7-X optimization \cite{Beidler1990,Grieger1992} presumably used similar profiles to obtain a $\beta$-limit of $\sim 5\%$ imposed by linear ideal ballooning instability. 
However, such a profile can be difficult to achieve in experiments. In fact, the pressure profiles achieved in W7-X experiments with higher performance so far tend to be more peaked \cite{Grulke2025,Langenberg2024}. 
Therefore, in this section, we consider a more peaked profile with $\bar{p}=(1-s)^2$ as shown by the red solid curve in Figure \ref{fig:growth}(b), and investigate how the results are affected. 

Here we still consider the standard EIM configuration, specifically two equilibria with $\beta=3.88\%$ and $4.04\%$, respectively. 
Their initial rotational transform profiles are shown by the red curves in Figure \ref{fig:prof}(a), and a notable difference from the broad-profile equilibrium (blue dashed) is the presence of $\iota=5/6$ resonances. 
These equilibria are found to be unstable to ideal ballooning modes by the COBRA code \cite{Sanchez2000}. 
For completeness, we show their linear growth rates in single-field-period M3D-$C^1$ simulations in Figure \ref{fig:growth}(a), which are lower than the broad-profile case discussed in Section \ref{sec:kappa}. 
Nevertheless, the corresponding saturated shapes of the pressure profile in Figure \ref{fig:prof}(b) show that the peaked profile is subject to more pronounced ballooning induced degradation even with lower $\beta$ and linear growth rates, suggesting that the nonlinear magnitude of the instability cannot be simply predicted from its linear growth. 
In particular, the profile change worsens significantly as $\beta$ increases slightly from $3.88\%$ and $4.04\%$, while in Ref.~\onlinecite{Zhou2024} the worsening is more modest as $\beta$ increases from $4.9\%$ to $5.4\%$. 
This suggests that with a peaked pressure profile in the standard W7-X configuration, the $\beta$-limit is not only lower but also more rigid. 
It also implies that in future W7-X experiments at higher $\beta$, core MHD activities may naturally broaden the pressure profile, similar to the temperature flattening seen in NSTX experiments and simulations \cite{Jardin2022}.

In Figure \ref{fig:snap}, we compare snapshots of the pressure change $\delta p$ from simulations using the broad [column (a)] and peaked [columns (b) and (c)] profiles. 
Row 1 shows coherent ballooning mode structures at the end of the linear growth, which are localized in the bad-curvature regions and radially coincide with where the pressure gradient is the largest. 
That is, the modes are located near the periphery and at mid-radius for the broad and peaked profiles, respectively. 
The saturated pressure changes in row 2 are located more radially inward compared to the linear modes, which is a feature that cannot be learned from linear analysis. 
\YZ{This implies that the saturation mechanism is unlikely a simple relaxation of profiles to linear stability, since the pressure gradient is still quite large where the linear modes are.}
In particular, the degradation of the peaked profile is concentrated in the core. 
Similar features have also been observed in simulations and measurements of ballooning-induced core collapses in the Large Helical Device \cite{Miura2001,Mizuguchi2009,Ohdachi2017,Sato2017,Varela2026,Civit-Bertran2025}.

In row 3 of Figure \ref{fig:snap}, we show the Poincar\'e plots of the saturated states in the simulations. 
In the simulation with the broad profile (a3), a large fraction of the magnetic field remains non-integrable in the periphery, in contrast to those with the peaked profile [(b3) and (c3)] where most of the flux surfaces have healed. This suggests that more integrable magnetic fields do not necessarily correspond to softer $\beta$-limits, and corroborates the finding in Ref.~\onlinecite{Zhou2024} that convection contributes more to the degradation than conduction. 
In the latter cases, $m=6$ islands have formed and can be naturally associated with the $\iota=5/6$ resonance shown in Figure \ref{fig:prof}(a). 
This is noteworthy since in Ref.~\onlinecite{Zhou2024}, one observation of the benign saturation is limited occurrence of low-$n$ activities, which may be owed to the absence of low-$n$ resonances in the rotational transform profile. 
In Section \ref{sec:iota}, we shall vary the magnetic configuration to see how the presence or absence of the $\iota=5/6$ resonance affects the nonlinear saturation of the ballooning modes.

\section{Influence of the rotational transform}\label{sec:iota}

\begin{figure}
\centering
\includegraphics[width=0.5\textwidth]{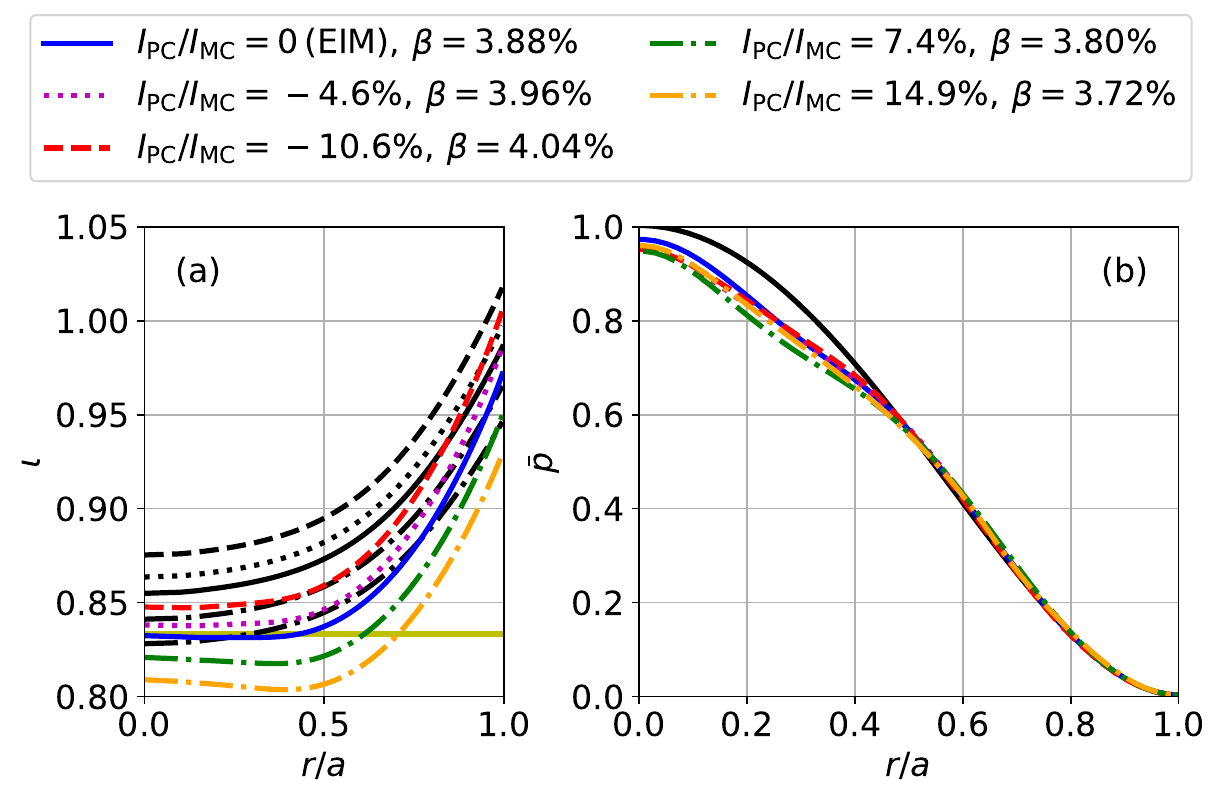}
\caption{(a) The rotational transform profiles in the simulated finite-$\beta$ equilibria (colored), with the vacuum profiles (black) shown for comparison. Different values of $I_\mathrm{PC}/I_\mathrm{MC}$ are distinguished by line styles. The yellow line marks the $\iota=5/6$ resonance. (b) The final shapes of the pressure profile in the M3D-$C^1$ simulations, with the initial shape (black solid) shown for comparison.}
\label{fig:iota}
\end{figure}

The W7-X coil system is designed with the flexibility to support a variety of magnetic configurations \cite{Geiger2015}. 
The standard EIM configuration is achieved when all five kinds of modular coils carry the same current, $I_\mathrm{MC}$. 
Then, the rotational transform can be changed by applying equal currents in the two kinds of planar coils, $I_\mathrm{PC}$. 
In this section, we adjust the ratio $I_\mathrm{PC}/I_\mathrm{MC}$ while keeping the core magnetic field and plasma volume roughly constant to generate a series of vacuum configurations, whose rotational transform profiles are shown by the black curves in Figure \ref{fig:iota}(a) and distinguished by line styles. 
The solid curve is the EIM configuration in Figure \ref{fig:prof}, and it can be seen that a positive or negative $I_\mathrm{PC}$ decreases or increases the rotational transform, respectively. 
Meanwhile, the shape of the profile and the magnetic shear are roughly maintained. 

\begin{figure*}
\centering
\includegraphics[width=\textwidth]{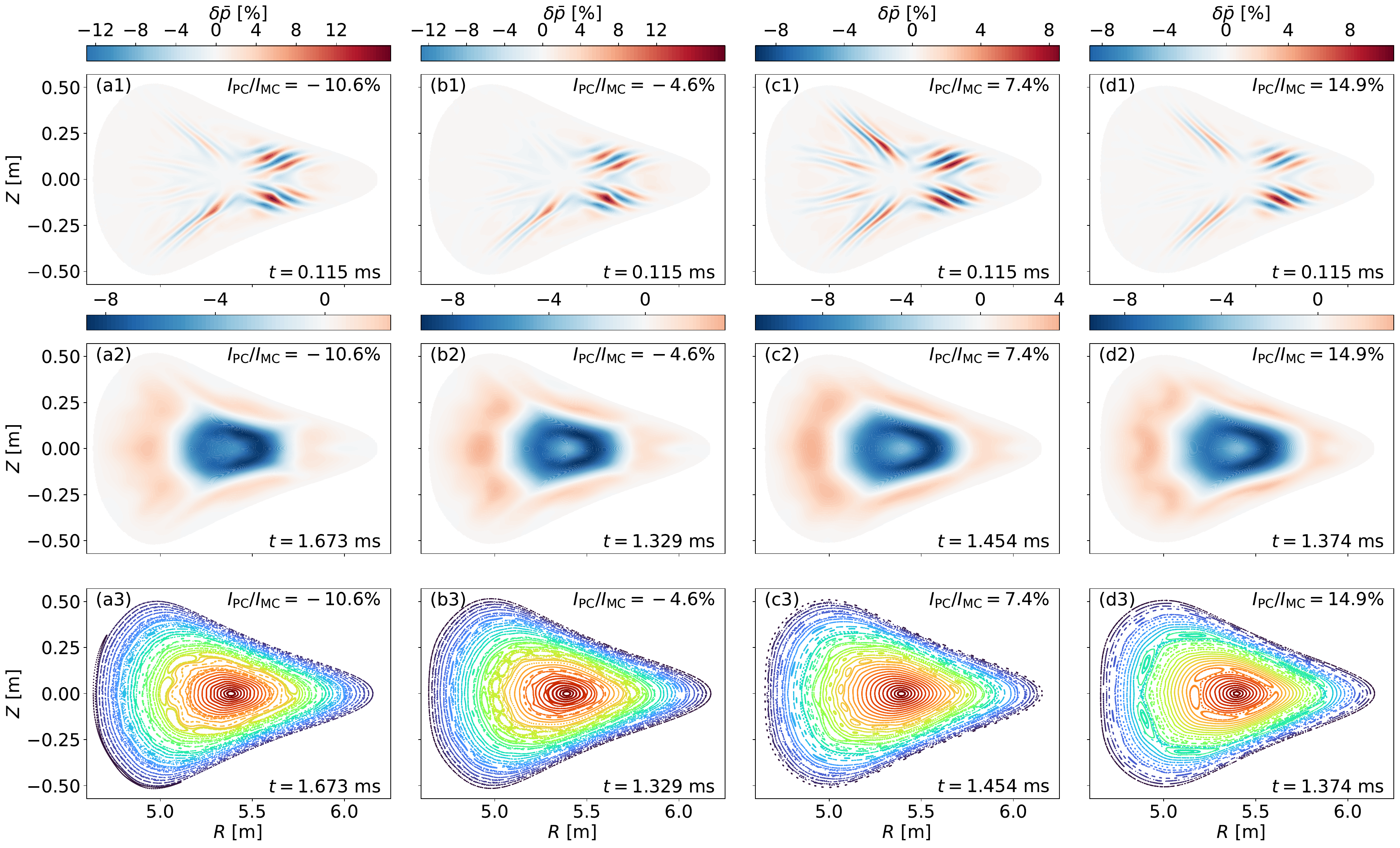}
\caption{Snapshots of the normalized pressure change in the simulations in Figure \ref{fig:iota} (except the EIM case): row 1 shows mode structures at the end of the linear growth phase, while row 2 shows the saturated states at the end of the simulations. Note the different scales in the color bars. Row 3 shows the Poincar\'e plots of the saturated states, where different colors label different field-lines.}
\label{fig:snap2}
\end{figure*}

For finite-$\beta$ equilibria, we use the peaked profile in Section  \ref{sec:prof} and adjust $\beta$ so that the linear growth rates are comparable ($\gamma\approx 0.07~\mu s^{-1}$).
The corresponding finite-$\beta$ rotational transform profiles are shown by the colored curves in Figure \ref{fig:iota}(a), where the blue solid curve corresponds to the red dotted curve in Figure \ref{fig:prof} and column (b) in Figure \ref{fig:snap}. 
Notably, an $\iota=5/6$ resonance is present for positive $I_\mathrm{PC}$ and absent for negative $I_\mathrm{PC}$.
In Figure \ref{fig:iota}(b), we present the saturated pressure profiles in full-torus simulations, which show similar levels of degradation despite the varying rotational transform and the presence or absence of the $\iota=5/6$ resonance. 
In row 1 of Figure \ref{fig:snap2}, the mode structures at the end of the linear growth phase show no obvious difference. 
The saturated states shown in row 2 are also largely similar, despite the different island structures in the corresponding Poincar\'e plots shown in row 3. 
The seeming invariance of the saturated state with regard to the change in rotational transform (and the presence or not of a low-order resonance) suggests that the saturation mechanism is not specific to a particular mode, resonant or non-resonant.
Therefore, it might be possible to formulate a reduced theory to predict the saturation amplitude or state that is agnostic of the detailed dynamics, for example, from an energetic perspective. Meanwhile, there are other key factors in the magnetic configuration, such as magnetic shear, that have not been varied in this work. Their impact will be studied in future work.

\section{Summary and discussion}\label{sec:sum}
In this paper, we present comprehensive MHD simulations of ideal ballooning modes in high-$\beta$ W7-X plasmas, building on our recent work that predicts benign saturation above the designed $\beta$-limit \cite{Zhou2024}.
First, we find that while increasing the parallel thermal conductivity reduces the linear growth rate, the impact on the saturated pressure profile is limited. 
Second, we show that an equilibrium with a peaked profile and lower $\beta$ is subject to more significant degradation than a broad profile with higher $\beta$ and a larger growth rate, suggesting that benign saturation, or nonlinear stability, is not guaranteed and not dictated by linear growth. 
Third, we vary the equilibrium rotational transform by adjusting the planar coil current, and find that similar growth rates result in similar magnitudes of profile change regardless of the presence of a low-order resonance, which implies that the saturation mechanism is not specific to a resonant or non-resonant mode. 
These results indicate that MHD stability should still be treated seriously in stellarator operation and design, for which nonlinear modeling using tools like M3D-$C^1$ can play an instrumental role. 

That said, a full-torus M3D-$C^1$ simulation typically takes hundreds of thousands of CPU hours, which is too costly for iterative optimization of operation scenarios or device designs. To incorporate nonlinear stability considerations in these applications, a reduced model for predicting the nonlinear saturation amplitude or state needs to be developed. The simulation results in Section \ref{sec:iota} imply such a possibility and can offer useful insights and help verify the reduced model. This goal shall be pursued in our future work.

\acknowledgments
This research is supported by the National MCF R\&D Program under Grant Number 2024YFE03230400, the National Natural Science Foundation of China under Grant Number 12305246, the Fundamental Research Funds for the Central Universities, the State Key Laboratory of Nuclear Physics and Technology (No. NPT NPT2025ZX03), and the U.S. Department of Energy under contract number DE-AC02-09CH11466. The United States Government retains a non-exclusive, paid-up, irrevocable, world-wide license to publish or reproduce the published form of this manuscript, or allow others to do so, for United States Government purposes.

\end{document}